\documentclass[prd,twocolumn,aps,preprintnumbers,letterpaper,showpacs]{revtex4}
\usepackage{amsmath}
\usepackage{tipa}
\usepackage{amssymb}
\usepackage{graphicx}
\usepackage{bm}
\usepackage{enumerate}
\usepackage{color}
\usepackage[colorlinks = true,
            linkcolor = blue,
            urlcolor  = blue,
            citecolor = blue,
            anchorcolor = blue]{hyperref}
\usepackage{subfigure}

\def\be{\begin{equation}}
\def\ee{\end{equation}}

\allowdisplaybreaks[2]

\newcommand{\bear}{\begin{align}}
\newcommand{\eear}{\end{align}}
\newcommand{\bea}{\begin{align}}
\newcommand{\eea}{\end{align}}
\newcommand{\nn}{\nonumber}

\def\hri#1#2{\href{http://arxiv.org/abs/#1}{[ArXiv:#1]#2}}
\def\hre#1#2{\href{http://arxiv.org/abs/#1/#2}{[ArXiv:#1/#2]}}

\newbox\pippobox

\def\II{\relax{\rm I\kern-.18em I}}

\def\l{\lambda}

\def\Awf{{A}}

\begin{document}

\title{Holographic entropy and real-time dynamics of quarkonium dissociation\\ 
in non-Abelian plasma}

\author{Ioannis Iatrakis$^{1, 3}$} 
\email[Email: ]{ioannis.iatrakis@stonybrook.edu}
\author{Dmitri E. Kharzeev$^{1, 2}$}
\email[Email: ]{dmitri.kharzeev@stonybrook.edu }
\affiliation{$^1$Department of Physics and Astronomy,
Stony Brook University,  Stony Brook, NY 11794-3800, USA \\
$^2$Department of Physics and RIKEN-BNL Research Center, Brookhaven National Laboratory, Upton New York 11973-5000, USA\\
$ ^{3}$Institute for Theoretical Physics, Utrecht University Leuvenlaan 4, 3584 CE Utrecht, The Netherlands}

\date{\today}

\begin{abstract}
 The peak of the heavy quark pair entropy at the deconfinement transition, observed in lattice QCD, suggests that the transition is effectively driven by the increase of the entropy of bound states. The growth of the entropy with the inter-quark distance leads to the emergent 
 entropic force that induces dissociation of quarkonium states. Since the quark-gluon plasma around the transition point is a strongly coupled system, we use the gauge-gravity duality to study the entropy of heavy quarkonium and the real-time dynamics of its dissociation. In particular, we employ the Improved Holographic QCD model as a dual description of large $N_c$ Yang Mills theory. Studying the dynamics of the fundamental string between the quarks placed on the boundary,  we find that the entropy peaks at the transition point. We also study the real-time dynamics of the system by considering the holographic string falling in the black hole horizon where it equilibrates. In the vicinity the deconfinement transition, the dissociation time is found to be less than a fermi, suggesting that the entropic destruction is the dominant dissociation mechanism in this temperature region.
\end{abstract}

\maketitle

\section{Introduction}

The heavy quarkonium is an important probe of the finite temperature QCD matter.  In particular, the dissociation of the quarkonium has been proposed as a signature of deconfinement, \cite{MaSa}. Lattice QCD results \cite{Kaczmarek:2005gi} indicate that the entropy of the heavy quark-antiquark pair has a sharp peak at the deconfinement transition. Recently it was proposed \cite{Kharzeev:2014pha,Hashimoto:2014fha} that this peak reflects the nature of deconfinement transition that may be driven by the entropy associated with the bound states of QCD. Moreover, the growth of the entropy with the inter-quark distance gives rise to the entropic force that drives the dissociation of heavy quarkonium \cite{Kharzeev:2014pha}. 

In the holographic approach, the peak of the entropy emerges when the string stretched between the heavy quarks touches the horizon of black hole \cite{Hashimoto:2014fha}. In terms of the boundary theory, this is likely associated with the condensation of ``long strings" spanning the entire volume of the finite temperature system \cite{Kogut:1974ag,Deo:1989bv,Hashimoto:2014xta}. 

In this paper we extend the holographic studies of \cite{Hashimoto:2014fha} by using the 
 the Improved Holographic QCD (IHQCD) model, \cite{ihqcd1, ihqcd2}. This is an Einstein-dilaton holographic model of large $N_c$ 4d Yang Mills theory that reproduces quite well its low energy behavior including the spectrum of hadrons and thermodynamics \cite{ihqcdT, data}. We compute the entropy, $S(T)$, of a heavy quark-antiquark pair as a function of temperature, $T$, and find a peak of $S(T)$ at the confinement-deconfinement transition temperature, $T_c$. Using the fit of the parameters that was made in \cite{data} and without introducing any other phenomenological parameters we find that $S(T)$ agrees with the lattice result for $T>1.1T_c$. However the peak at $T=T_c$ is lower than the peak observed on the lattice, and the high temperature asymptotics of $S(T)$ differs from the lattice QCD result. This latter discrepancy appears in the UV where we do not expect the classical treatment of the gravity side valid. However, it is possible that some modification of the used Nambu-Goto action may be done to reproduce the lattice result better in the vicinity of $T_c$. 
 
We also study the real-time dynamics of quarkonium dissociation in the QCD medium. Holographically, the background geometry is a black hole space-time describing a certain temperature. The quark-antiquark pair is placed on the boundary with inter-quark distance $L$ with a string that stretches between them.  Initially the string lies on the boundary, and then falls towards the black hole horizon under the gravitational force of the background metric. When the string reaches the horizon the system reaches its equilibrium state in which the string is split into two pieces. Each piece of the string has an endpoint on the boundary (on the quark or the antiquark) and the string stretches along the holographic coordinate and falls inside the black hole horizon. This state corresponds to a deconfined phase where the quarkonium is dissociated. Our numerical result suggests that the quarkonium dissociation is fast -- less than one fermi around $T=T_c$. The holographic string as a model of the quarkonium has also been previously studied in other holographic theories in order to describe thermodynamics and thermalization, see \cite{Lin:2006rf, Chesler:2008uy, Finazzo:2014zga}.

\section{IHQCD}

The Improved holographic QCD is an effective five dimensional holographic model of (3+1) dimensional $SU(N_c)$ Yang Mills (YM) theory \cite{ihqcd1}. The complete action of the IHQCD model is given by
\be
S_g= M_p^3 N_c^2 \int d^5x \ \sqrt{-g}\left( R- {4 \over 3}{
(\partial \lambda)^2 \over \lambda^2} + V_g(\lambda) \right) +S_{HG}\, ,
\label{glueact}
\ee
where $S_{HG}$ is the Hawking-Gibbons term and $M_p$ is the 5-dimensional Planck mass. The real scalar field $\l=e^\phi$ (where $\phi$ is the the dilaton field) is dual to the ${\mathbb Tr} F^2$    operator, and is identified as the holographic 't Hooft coupling; $V_g(\lambda)$ is the dilaton potential. The Ansatz for the vacuum solution of the metric is
\be
ds^2=e^{2 \Awf(r)} \left( -f(r)\,dt^2+dx_{3}^2+{dr^2 \over f(r)} \right) \,,
\label{bame}
\ee
where $dx_{3}$ denotes the spatial line element, the warp factor $A$ is identified as the logarithm of the energy scale in the field theory and $f(r)$ is the black hole factor.
The factor $f(r)$ is equal to 1 in the confined phase of the theory and is a non-trivial function of $r$ in the deconfined phase. The position of the black hole horizon, $r_h$, is identified by $f(r_h)=0$.  

Our convention will be that the UV boundary lies at $r=0$, and the bulk coordinate therefore runs from zero to infinity or to the black hole horizon, depending on the phase of the theory. In the UV, $r$ represents the inverse of the energy scale of the dual field theory. The near boundary asymptotics of the model are such that they match the perturbative expansion of YM. The metric approaches the AdS geometry and the dilaton vanishes logarithmically to model the running of the YM coupling. 

The dilaton potential has an analytic expansion in terms of $\lambda$ as $r\to 0$ (and $\l \to 0$), $ V_g(\lambda)={12/ \ell^2} \left( 1+ v_0 \lambda + \ldots \right)$, where $\ell$ is the AdS radius and the coefficients are matched to the YM $\beta$-function.  Then, the metric and the dilaton close to the boundary read 
 
 \be
 A \sim -\log\left({r \over \ell}\right)+ {4\over 9\log(\Lambda r)}+\dots \,\, , \,\,\,  \lambda\sim -{8 \over 9 v_0 \log(\Lambda r)}+\ldots\, ,
 \label{uvas}
 \ee
 where $\Lambda$ corresponds to the $\Lambda_{QCD}$ scale.  The UV structure of the model matches the asymptotic perturbative expansion of YM -- however the model is not expected to describe the UV limit of the theory since the field theory is weakly coupled and its holographic dual is expected to be a string theory in this region. The current asymptotics provide a reliable choice of boundary conditions of the model. Nevertheless one has to keep in mind that IHQCD is an effective approach to low energy YM theory, and hence its IR (or large $r$) structure is more relevant for our discussion.
 
The IR limit of the theory is strongly coupled, hence we expect the dilaton in the confined phase to diverge in the IR. The large $\lambda$ expansion of the potential determines the IR physics of the IHQCD model. To reproduce confinement and to ensure that the glueball spectrum is gapped, discrete and follows the linear Regge trajectories asymptotically, one chooses the IR asymptotics of the dilaton potential as $V_{g}(\l \to \infty) \sim \l^{4\over 3} \sqrt{\log(\l)}$ \, . The solution for the metric and the dilaton is

\be
A(r) \sim -r^2 \,\,,\,\,\, \l(r) \sim r e^{{3\over 2}r^2}\,.
\ee
In the deconfined phase, $A$ and $\l$ have a regular expansion around the horizon and $f$ vanishes. A simple interpolation of the UV and IR asymptotics leads to the choice of the dilaton potetnial
 \be
 V_g(\lambda)={12 \over \ell^2} \left[1+V_0 \l +V_1 \l^{4/3} \sqrt{\log (1+V_2 \l^{4/3} +V_3 )} \right]\,. 
  \ee 
Upon matching to the perturbative running of the Yang-Mills coupling, two out of the four of the parameters $V_0, V_1, V_3$ and $V_4$ are left independent.  Those are fixed by matching to lattice results for two thermodynamic quantities, the latent heat and the entropy density at the deconfinement transition. Then, the model describes successfully the zero-T glueball spectrum and the thermodynamics above the confinement-deconfinement transition, \cite{data}. The five dimensional Planck mass $M_p$ is determined by requiring that the high temperature asymptotics of the free energy follows Stefan-Boltzman law. This fixes 

\be
(M_p \ell)^3 ={1\over 45 \pi^2} \,.
\ee

\section{The quark-antiquark pair}

The interaction of a pair of heavy quark and antiquark in the boundary field theory is modeled holographically by the dynamics of a Nambu-Goto string in the bulk, \cite{Maldacena:1998im}. The quark and antiquark, which are located at distance L at the boundary, are attached at the endpoints of the string which extends into the bulk.
The Nambu-Goto action reads

\be S_{NG}=-T_f \int d\tau d\sigma \sqrt{- \det  (g_S)_{\mu\nu} \partial_\alpha X^\mu \partial_\beta X^\nu } \, , \ee  
where the string frame metric is used and $T_f$ is the string tension. The string-frame metric is related to the Einstein metric by

\be 
(g_S)_{\mu\nu}=e^{2A_s(z)}\eta_{\mu\nu}\,\,,\,\,\, A_s(z)=A(z)+{2\over 3} \Phi(z)  \, .
\ee
The free-energy of the quark-antiquark pair is equal to the on-shell  Nambu-Goto action

\be
T F(L)= S_{NG}[X_{min}] \, ,
\ee
where $X_{min}^{M}$ is the solution for the embedding of the string in the background with the minimum area.

\subsection{Confined phase}

The string profile in the confined phase of IHQCD model was studied in \cite{ihqcd2}; let us briefly review it here .The world sheet coordinates are taken to be $\tau=t$ and $\sigma=x$ and the string is embedded in the 5D bulk space-time, i.e. $r=r(x)$.

\be
S_{NG}=T_f \int_0^L dx \, e^{2 A_s} \sqrt{1+r'(x)^2} \, .
\ee
 In the confined phase of the theory, the world-sheet has a turning point $r_*$ determined by the condition $r'(x)=0$.  The quark-antiquark distance $L$ is expressed in terms of $r_*$ as
 
 \be
 L=2 \int_{\epsilon}^{r_*} {dr \over \sqrt{e^{4 A_s(r)-A_s(r_*)} -1}}
 \ee
 and the free energy is 
 
 \be
 F(L)=-2 T_f \int_{\epsilon}^{r_*} dr{e^{4 A_s(r)} \over \sqrt{e^{4 A_s(r)-4 A_s(r_*)}-1}} \,.
\ee
The point $r_*$ is a minimum of $A_s(r)$. In the limit of large $L$ we find the potential energy of the quark-antiquark pair is given by

\be
F(L) = T_f e^{2 A_s(r_*)} L + \ldots \, ,
\ee
which is a linear law signaling confinement. In \cite{data}, the Nambu-Goto string tension $T_f$ was determined by matching to the confining string tension as found by lattice methods. Their relation is $\sigma=T_f e^{2 A_s(r_*)}$. The fit to the data results in 

\be
T_f \ell^2 = 6.5 \, ,
\ee
where the tension is measured in units of the AdS radius.

\subsection{Deconfined phase}

\begin{figure}[!tb]
\begin{center}
\includegraphics[width=0.49\textwidth]{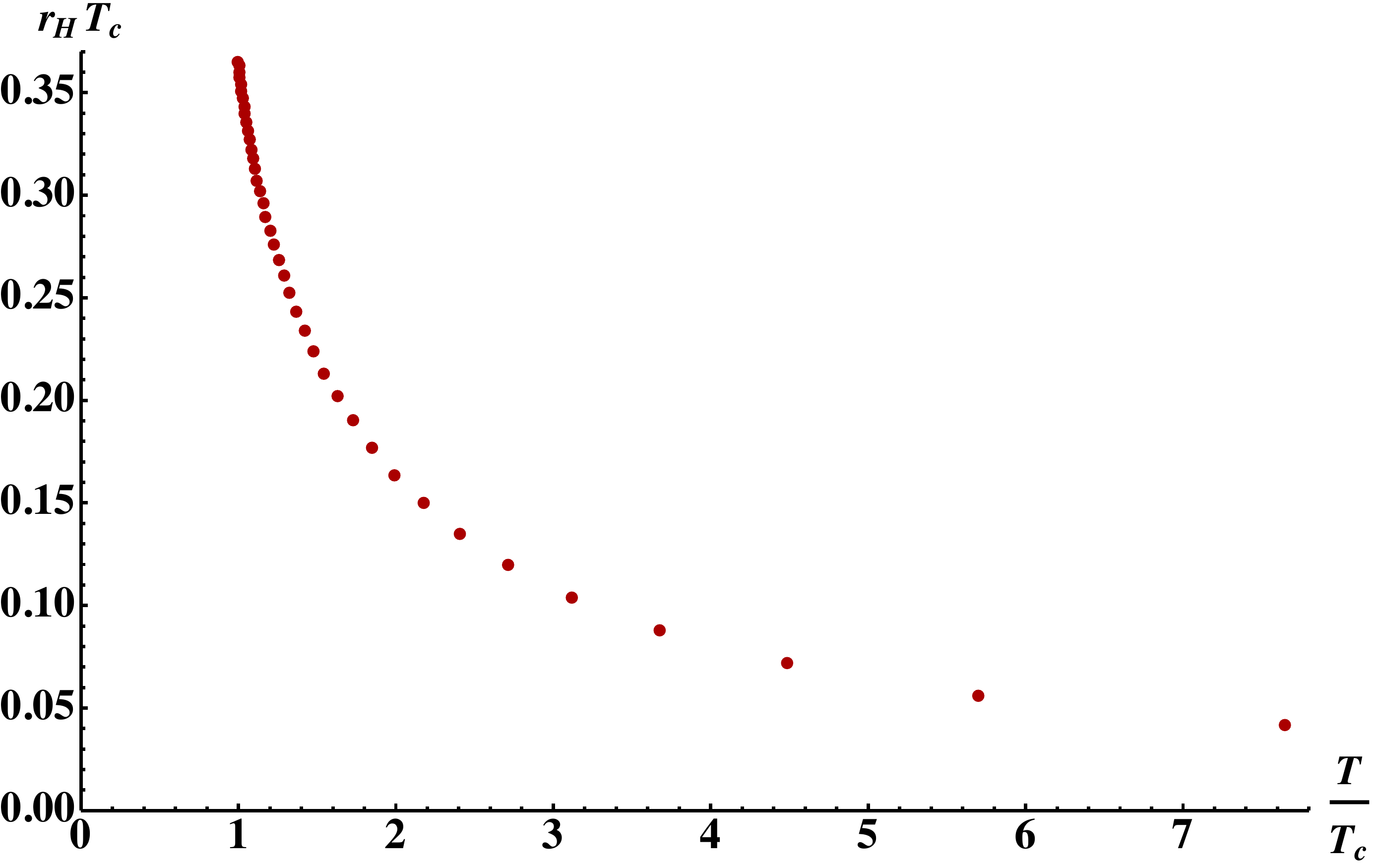}
\caption{The horizon of the large black hole branch in terms of the temperature, in units of critical temperature.}
\label{zht}
\end{center}
\end{figure}

The holographic dual of the deconfined phase of the field theory is 	a black hole metric in the bulk. In this case the string falls into the horizon, so its embedding in the bulk is a straight line starting from the boundary and extending into the horizon, $X^{M} =(t,x,0,0,r) $. The free energy is equal to the on-shell string action

\be
F(L)=- 2 T_f \int_{\epsilon}^{r_H} dr \, e^{2 A_s(r)} \, .
\ee
This free energy has been calculated for a gravity-dilaton holographic model in the same class as IHQCD in \cite{Finazzo:2014zga}. Then, the entropy of the quark-antiquark pair follows 

\be
S=-{\partial F \over \partial T}=2 T_f e^{2 A_S(r_H)} {\partial r_H \over \partial T} \, .
\label{entads}
\ee
The temperature of the black hole is defined as $T= |f'(r_H)|/(4 \pi)$. In \cite{ihqcd3}, the phase structure of IHQCD was studied.  It was shown that confining gravity-dilaton theories admit
 black hole solutions above a minimum value of temperature, $T_{min}$. For $T>T_{min}$, two branches of black hole solutions for a given value of the temperature exist. Typically,  one branch has a large horizon which is closer to the boundary (i.e. $z_H$ is small) and the other branch has a small horizon (large $z_H$). The large black hole branch is thermodynamically stable, while the small branch is unstable. Hence the large black hole branch is dual to the deconfined phase of the field theory. Moreover, the model is shown to exhibit a first order confinement-deconfinement phase transition at some critical temperature, $T_c$. This is a direct analogue of the Hawking-Page transition which was found for Einstein-Hilbert action with a cosmological constant.   In Fig. \ref{zht}, we show the position of the horizon in terms of the temperature in units of $T_c$.

The high-T asymptotics of the entropy as calculated by perturbation theory in Yang-Mills is \cite{Kaczmarek:2005gi},

\be
S_{YM} = {8^{3/2} \pi^2 \over 3 \,11^{3/2}}  {1 \over \log^{3\over 2} \left( {\pi \,T \over \Lambda } \right) } \, .
\ee

From the AdS point of view, the high-T asymptotics of (\ref{entads}) corresponds to a large black hole with the horizon located in the the near-boundary region of the bulk space-time. When the horizon is located close to the boundary the temperature is given by the pure AdS formula

\be 
T={1 \over \pi \, r_H} \, ,
\ee
where we have taken $f(r)=1-{r^4 \over r_H^4}$. Using the UV expansions of the bulk fields  $A(r)$ and $\l(r)$ given in Eq(\ref{uvas}), Eq.(\ref{entads}) becomes

\be
S_{IHQCD} \simeq {2 T_f \ell^2 } \left( {8 \over 9 V_1}\right)^{4/3} {1 \over \log^{4 \over 3} \left( {\pi T \over \Lambda} \right) }
\ee 
The power of the logarithm in this ``Nambu-Goto" entropy at high T is slightly higher than the perturbative result. This means that the UV asymptotics of the Nambu-Goto action does not seem to match the perturbative QCD, even though the power of the logarithm is numerically close in the two formulae.  

\begin{figure}[!tb]
\begin{center}
\includegraphics[width=0.49\textwidth]{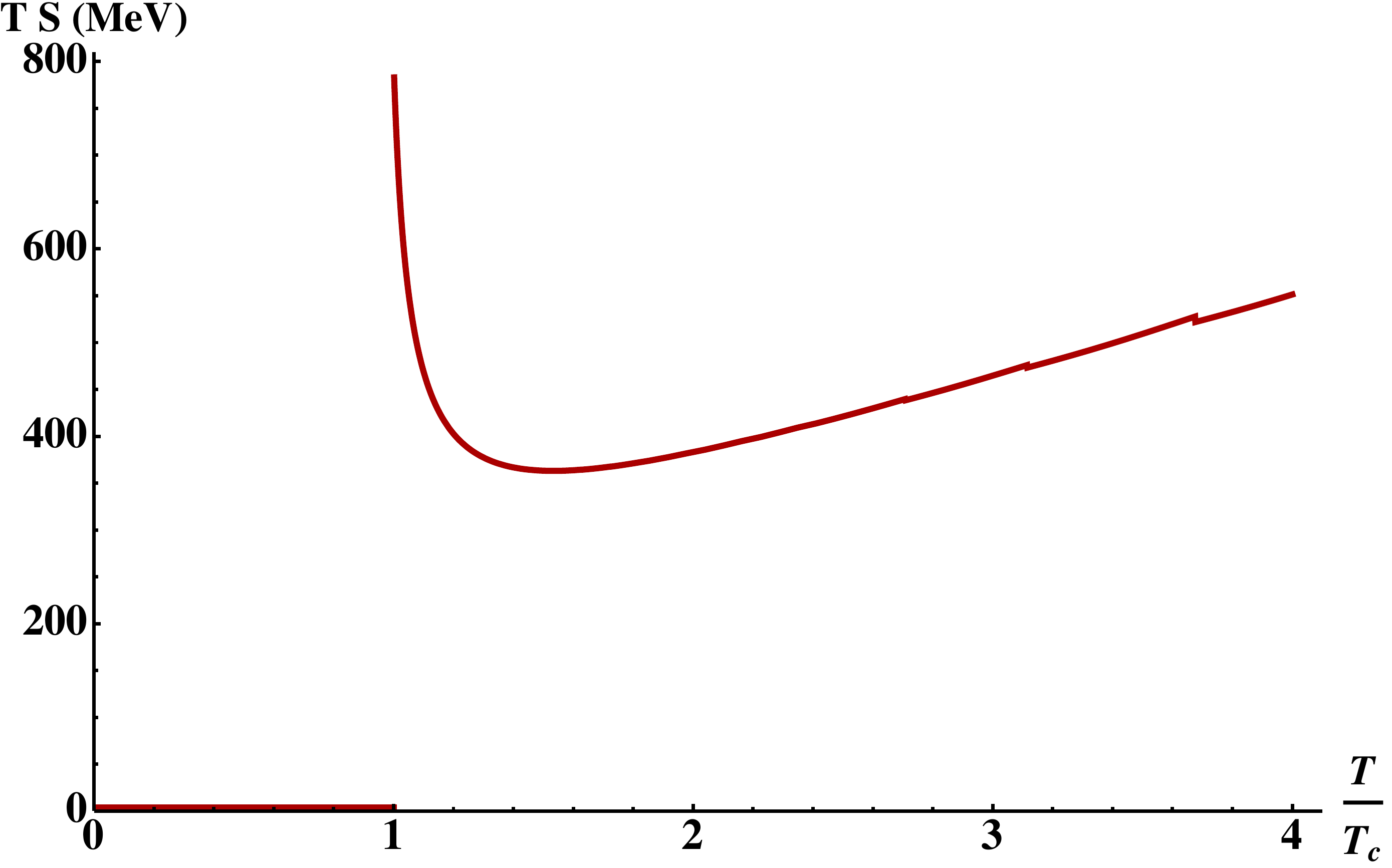}
\caption{The entropy of the heavy quark-antiquark pair in the deconfined phase.}
\label{entropy1}
\end{center}
\end{figure}

In Fig. (\ref{entropy1}), we show the entropy of the quark-antiquark pair as a function of $T/T_c$ for the IHQCD model. In the confined phase, the background does not depend on temperature since the bulk theory describes the large $N_c$ limit of the field theory. In the deconfined phase, the bulk solution is a dilaton black hole that results in an entropy which notably is in qualitative agreement with the lattice result, \cite{Kaczmarek:2005gi}. It should be noticed that in the calculation of the quark-antiquark entropy we have not changed the parameters of the model from the original fit in \cite{data}. At temperatures $T>1.1 T_c$, our results agree with the lattice data quantitatively.

\section{The real-time dynamics of quarkonium dissociation}

We now address the question of the quarkonium thermalization. We will assume that a pair of heavy quark and antiquark is produced at a fixed relative distance $L$ at the boundary.  If the quarks are sufficiently far apart at a given temperature, the gluon cloud around the pair will eventually thermalize and become part of the medium.  At this point in time the heavy quarkonium will be dissociated. 

The holographic description of the dissociated quark pair  is given by a string with its two endpoints fixed at the boundary and the string extending towards the black hole horizon. The string falls in the background gravitational field and eventually reaches the horizon. Then it equilibrates becoming a straight string falling in the black hole. The string coordinate is then $X^{M}=(t,x,0,0,r(t,x))$. The Nambu-Goto action then reads

\be
S_{NG}=-T_f \int dtdx e^{2A(r)}  \sqrt{f(r)-{\dot{r}^2 \over f(r)}+r'^2 } \, ,
\ee
where $r=r(t,x)$, $\dot{r}=\partial_t r$ and $r'=\partial_x r$. The equation of motion for the string is then 

\begin{align}
&\partial_t \left( {e^{2 A_s(r)} \dot{r}  \over   f(r)  \sqrt{f(r)-{\dot{r}^2 \over f(r)}+r'^2} } \right) 
-\partial_x \left( {e^{2A_s(r)} r'  \over   \sqrt{f(r)-{\dot{r}^2 \over f(r)}+r'^2} } \right) \nn \\
 & -\partial_r \left(e^{2 A_s(r)} \right) \sqrt{f(r)-{\dot{r}^2 \over f(r)}+r'^2} \nn \\ 
& - {e^{2 A_s(r)} \partial_{r} f(r) \over 2 \sqrt{f(r)-{\dot{r}^2 \over f(r)}+r'^2}} \left( 1+{\dot{r}^2 \over f(r)^2} \right)=0 \, ,
\label{streqfull}
\end{align}
\begin{figure}[!tb]
\begin{center}
\includegraphics[width=0.49\textwidth]{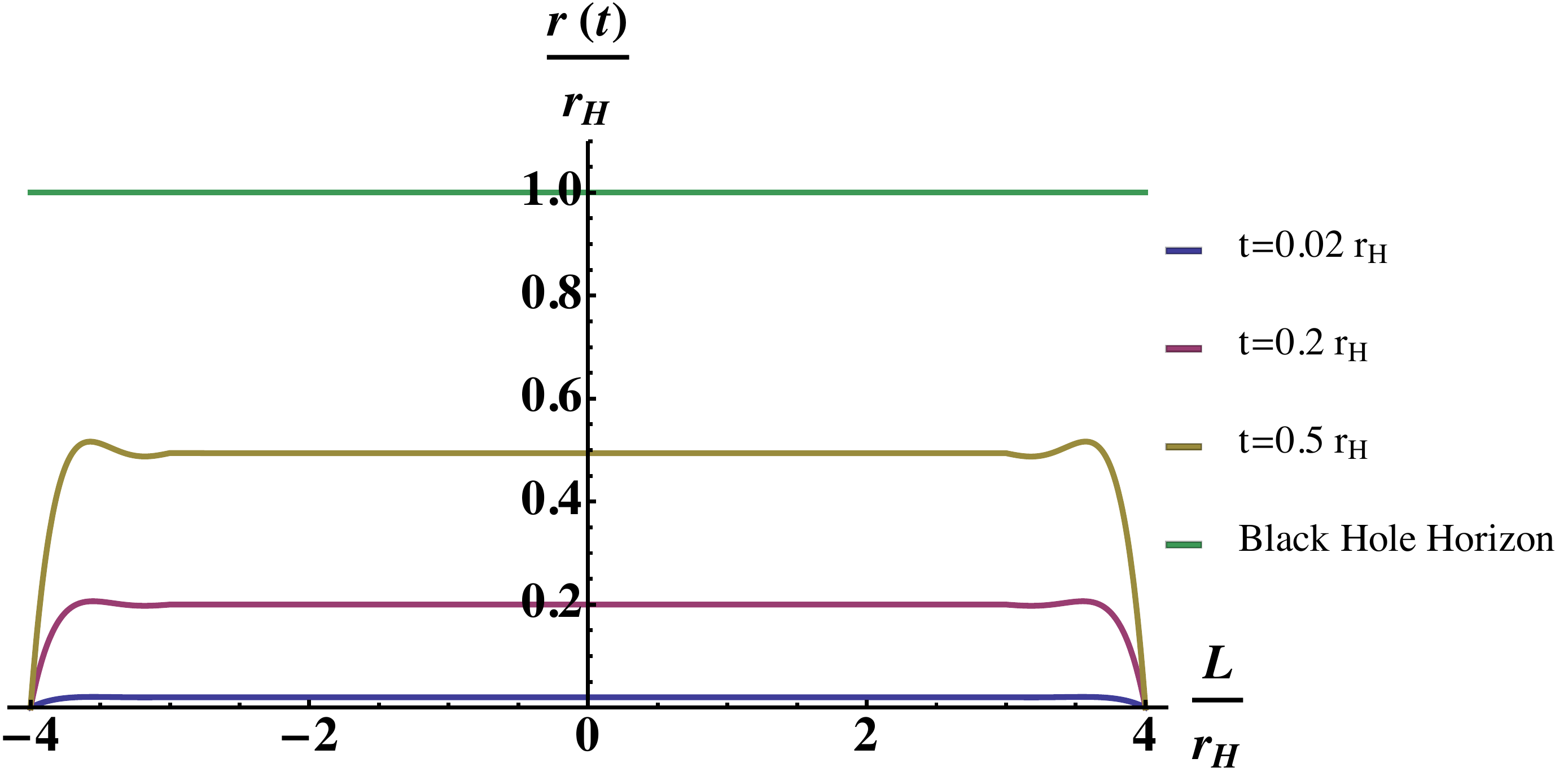}
\caption{The string extending towards the horizon, for $L= 8 r_H$, in case of the AdS-Schwarzschild black hole.}
\label{stringmotads}
\end{center}
\end{figure}
We numerically solve Eq. \ref{streqfull} in the case of AdS-Schwarzschild black hole, where $\lambda(r)=0$, $A(r)=\log\left( {\ell\over r}\right)$ and $f(r)=1-{r^4/r_H^4}$.  We consider the ends of the string to be fixed at certain distance on the boundary, $r(t,x=\pm L)=\epsilon$. The string is initially on the boundary $r(t=0,x)=\epsilon$ with zero velocity, $\dot{r}(t=0,x)=\epsilon$, where $\epsilon$ is the boundary cut-off. In Fig. \ref{stringmotads}, we show the profile of the falling string. Notice that  in case of large inter-quark distance $L$ the profile of the string is independent of $x$ and hence can be approximated by a straight string falling  towards the black hole horizon.  This means that $r=r(t)$, and the $x$ dependence is negligible for the bulk part of the string. In this simpler case, the Nambu-Goto action takes the form $ S_{NG}=-T_f \int dtdx\, e^{2A_s} \sqrt{ (f-{\dot{r}^2 \over f}) }$. 

In the case of small  distances $L$, the partial differential equation (\ref{streqfull}) more difficult to solve due to numerical errors. This is because the used coordinate system is singular at the horizon, hence one has to use Kruskal coordinates in order to solve the string equation of motion close to the horizon. We leave this problem for a future investigation.
\begin{figure}[!tb]
{\includegraphics[width=0.48\textwidth]{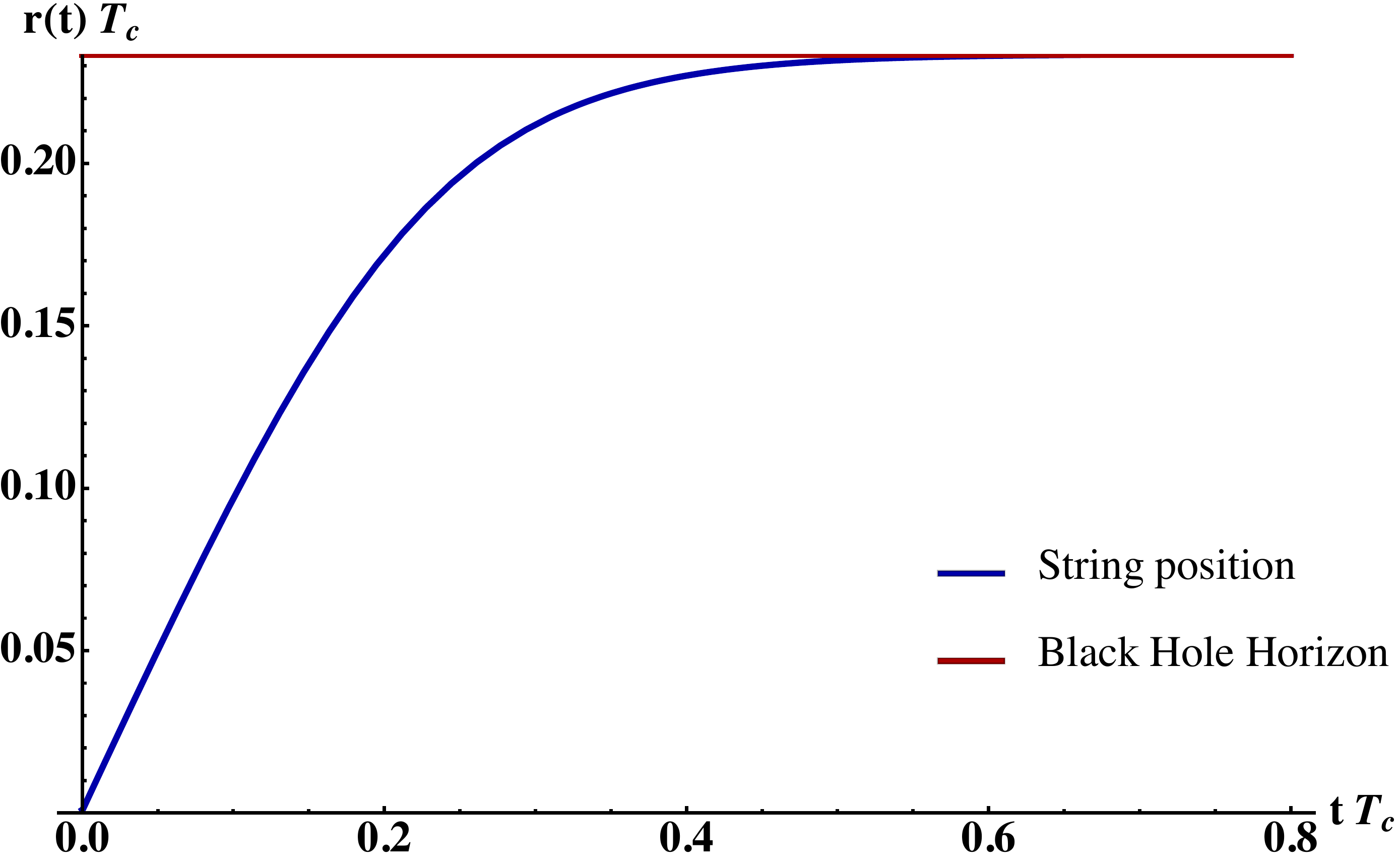},
\includegraphics[width=0.48\textwidth]{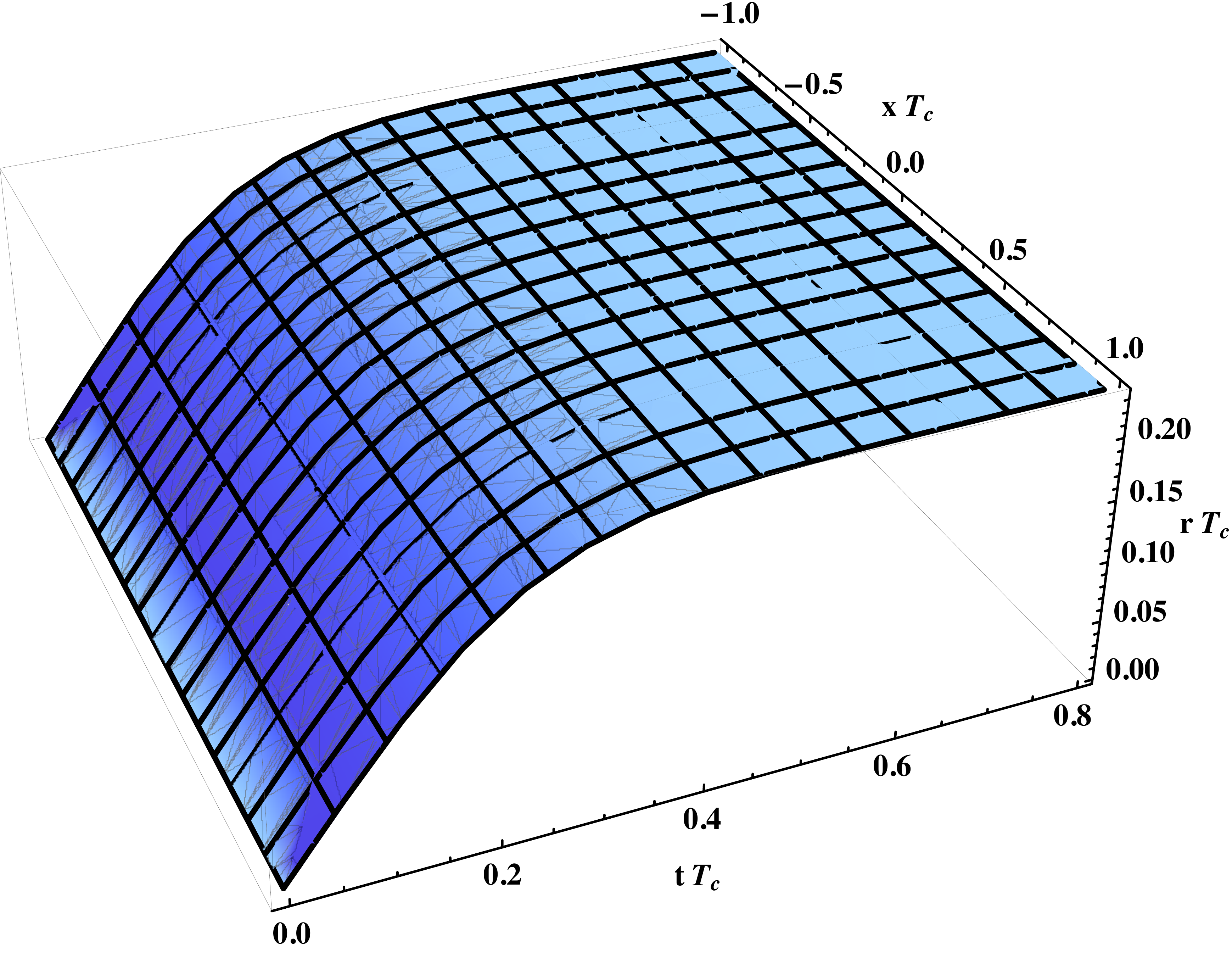}}
\caption{The string motion from the boundary to the horizon. The background black hole has temperature, $T=T_c$.}
\label{stringmot1}
\end{figure}

Coming back to the case of large inter-quark distances, the equation of motion of the string reads
\begin{align}
&\partial_t \left( {e^{2 A_s(r)} \dot{r}  \over   f(r)  \sqrt{f(r)-{\dot{r}^2 \over f(r)}} } \right)  -\partial_r \left(e^{2 A_s(r)} \right) \sqrt{f(r)-{\dot{r}^2 \over f(r)}} \nn \\
& - {e^{2 A_s(r)} \partial_{r} f(r) \over 2 \sqrt{f(r)-{\dot{r}^2 \over f(r)}}} \left( 1+{\dot{r}^2 \over f(r)^2} \right)=0 \, .
\end{align}
Since the Lagrangian does not explicitly depend on time, the energy is conserved

\be
E={T_f e^{2A_s} f \over \sqrt{f- {\dot{r}^2 \over f}}}\,,
\ee
and the velocity of the string is $ \dot{r}= {f \over E} \sqrt{E^2-T_f^2 e^{4 A_s} f} $. The string starts falling from the boundary of the bulk spacetime with initial velocity $\dot{r}(t=0)=0$. We may fix $E$ using this initial condition. The boundary of spacetime is set at cut-off distance $r=\epsilon$ which is determined by the initial energy of the created quark pair. Then, we easily determine the energy of the falling string as $E=T_f e^{2 A_s (\epsilon)}$.
Using the near-horizon asymptotics of the background fields,

\begin{align}
& A=A_h+A_1 (r-r_H)+\ldots \, , \,\, \lambda=\lambda_h+\lambda_1 (r-r_h)+\ldots \, , \nn \\ 
& f=-4 \pi T (r-r_H) \, ,
\end{align} 
we find  that the string approaches the horizon exponentially fast
\be
r(t)-r_H \simeq e^{-4 \pi T \, t} \,.
\ee
Solving numerically the full string equation of motion, we describe the entire motion from the boundary to the horizon. In Figure \ref{stringmot1}, we show the string position as a function  of time in units of $T_c$. When the string is close to the boundary it rapidly accelerates, and then asymptotically approaches the horizon. 

\begin{figure}[!tb]
\begin{center}
\includegraphics[width=0.49\textwidth]{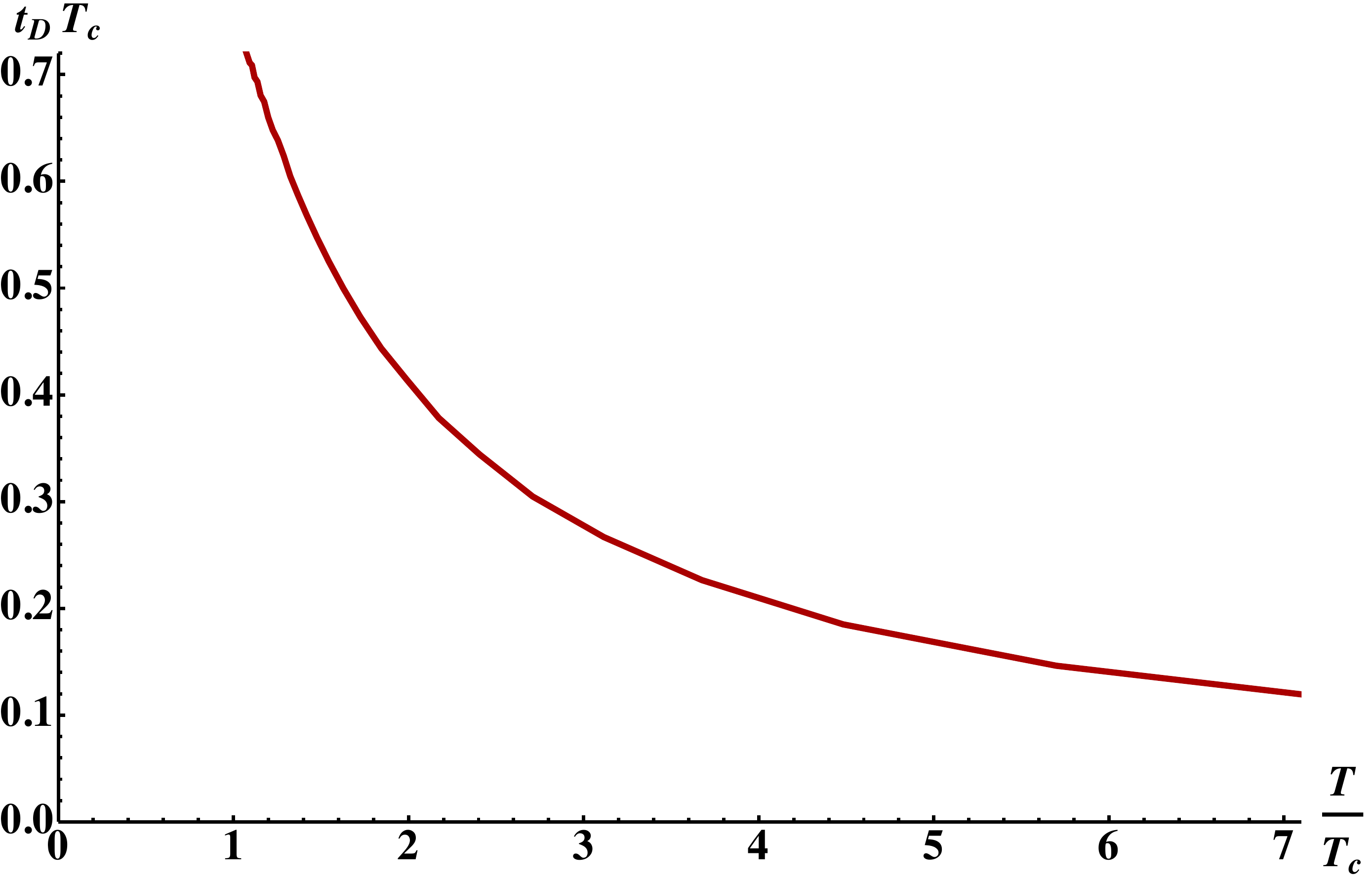}
\caption{The quarkonium dissociation time in terms of the temperature in units of $T_c$.}
\label{stringmot2}
\end{center}
\end{figure}

We can now calculate the time needed for the dissociation of the quarkonium -- it corresponds to the time needed for the string to reach the horizon from the boundary. Hence  we solve the string equation of motion for different black hole backgrounds, corresponding to states of different temperature, and compute the dissociation time, $t_D$. In Fig. \ref{stringmot2}, we plot  the dissociation time of the quark-antiquark pair in units of $T_c$ as a function of temperature. We see that the time is shorter for higher temperatures. This is natural, since the black hole is larger, and the string falls faster inside the horizon. Remarkably, the dissociation time of the heavy quarkonium around $T=T_c$  is less than a fermi -- therefore it is a fast process.

\section{Summary and Discussion}

The dissociation of quarkonium in the QGP conveys important information about the thermal quark-gluon medium and the onset of deconfinement. The peak in the entropy of the heavy quark pair observed on the lattice suggests an interesting picture of the deconfinement transition -- the destruction of the bound hadron states at the onset of deconfinement may be driven by the emergent entropic force \cite{Kharzeev:2014pha,Hashimoto:2014fha}. 

Here we have confirmed that the peak of the quark pair entropy at deconfinement is a generic feature of the Einstein-dilaton holographic models. 
Indeed,  in the confinement phase the entropy of the pair is zero since no temperature effects are seen in the large $N_c$ limit. This is not so in the deconfined phase, and Eq. (\ref{entads}) indicates that the entropy has a maximum at $T_c$ for the general class of black holes for which the position of the horizon is an increasingly rapid function of temperature near $T_c$, see Fig. \ref{zht}. In terms of the boundary theory, the entropy peak may be related to the condensation of QCD strings which leads to the formation of the QGP, see \cite{Kogut:1974ag,Deo:1989bv,Hashimoto:2014xta}. It will be very interesting to study this further using both holographic and lattice methods. In the latter case, the presence of long strings may be signaled by the unusual dependence of heavy quark pair observables on the lattice size -- naively, the color field of the pair does not extend beyond the Debye screening radius, but the long string spans the entire volume of the lattice. 

We have also presented the study of the real-time dependence of quarkonum dissociation. Our picture of dissociation corresponds to the string falling from the boundary to the horizon.  We have calculated the dissociation time for large (compared to 1/T) inter-quark distances and found that the thermalization process in IHQCD is very fast, with dissociation time less than a fermi at $T\sim T_c$. To study the case of finite quark-antiquark distance
one has to do a more refined numerical computation. This is an interesting problem, since 
solving it would provide an information about dissociation of quarkonium states of different size. We will present this analysis in a future work.

\vskip 1cm
{\bf Acknowledgements.} We thank Koji Hashimoto for useful discussions. 
This work was supported in part by the U.S. Department of Energy under Contracts No. DE-FG-88ER40388 and DE-SC0012704. This work
is also part of the D-ITP consortium, a program of the Netherlands Organisation for Scientific
Research (NWO) that is funded by the Dutch Ministry of Education, Culture and Science
(OCW).

\end{document}